\documentclass[journa,12pt,draftclsnofoot,onecolumn]{IEEEtran}
\usepackage[nocompress]{cite}
\usepackage[ruled,vlined]{algorithm2e}
\usepackage{times,amssymb}
\usepackage{amsmath}
\usepackage{subfigure}
\usepackage{bm}
\usepackage[font=small]{caption}
\usepackage[dvips]{graphicx}
\usepackage{url}
\newcommand{\subparagraph}{}
\usepackage{amsthm}
\usepackage{multirow}
\usepackage{color}

\ifodd 0
\newcommand{\rev}[1]{{\color{blue}#1}} %revise of the text
\newcommand{\com}[1]{\textbf{\color{red} (COMMENT: #1) }} %comment of the text
\newcommand{\comg}[1]{\textbf{\color{green} (COMMENT: #1)}}
\newcommand{\response}[1]{\textbf{\color{green} (RESPONSE: #1)}} %response to comment
\else
\newcommand{\rev}[1]{#1}

\newcommand{\com}[1]{}
\newcommand{\comg}[1]{}
\newcommand{\response}[1]{}
\fi
\begin{document}

\title{Exploiting Massive D2D Collaboration for Energy-Efficient Mobile Edge Computing}

%\markboth{Journal of \LaTeX\ Class Files}%
%{Shell \MakeLowercase{\textit{et al.}}: Bare Advanced Demo of IEEEtran.cls for IEEE Computer Society Journals}

\author{
        Xu~Chen,
        Lingjun~Pu,
        Lin Gao,
        Weigang Wu,
        and Di Wu

\IEEEcompsocitemizethanks{
\IEEEcompsocthanksitem Xu Chen, Weigang Wu and Di Wu are with School of Data and Computer Science, Sun Yat-sen University, Guangzhou, China. E-mails: chenxu35@mail.sysu.edu.cn,wuweig@mail.sysu.edu.cn, and wudi27@mail.sysu.edu.cn.
\IEEEcompsocthanksitem Lingjun Pu is with College of Computer and Control Engineering, Nankai University, Tianjin, China. E-mail: pulingjun@mail.nankai.edu.cn.
\IEEEcompsocthanksitem Lin Gao is with School of Electronic and Information Engineering, Shenzhen Graduate School, Harbin Institute of Technology, Shenzhen, China. E-mail: gaolin021@gmail.com.

%This work was supported in part by the National Key Research and Development Program of China under Grant 2016YFB0201900, National Natural Science Foundation of China (No. 61379157), and the Start-Up Fund from Sun Yat-sen University.
}
}

\maketitle

\begin{abstract}
In this article we propose a novel Device-to-Device (D2D) Crowd framework for 5G mobile edge computing, where a massive crowd of devices at the network edge leverage the network-assisted D2D collaboration for  computation and communication resource sharing among each other. A key objective of this framework is to achieve energy-efficient collaborative task executions at network-edge for mobile users. Specifically, we first introduce the D2D Crowd system model in details, and then formulate the energy-efficient D2D Crowd task assignment problem by taking into account the necessary constraints. We next propose a graph matching based optimal task assignment policy, and further evaluate its performance through extensive numerical study, which shows a superior performance of more than 50\% energy consumption reduction over the case of local task executions. Finally, we also discuss the directions of extending the D2D Crowd framework by taking into variety of application factors.
\end{abstract}

\section{Introduction}\label{sec:introduction}
More and more mobile applications such as mobile object recognition, IoT data stream processing, augmented reality and mobile health computing are emerging and will become prevalent in the 5G Era. These novel applications typically demand intensive computation resource for realtime processing and high network bandwidth for data exchange, leading to high energy consumption especially when they suffer from limited device resources \cite{schulman2010bartendr}. In general, mobile devices have limited energy capacity due to the physical size constraint, and the battery technology trend makes the limitation unlikely disappear in the near future \cite{guo2016optimal}. Therefore, it is of great significance to achieve energy-efficient task executions in mobile devices.

As an interesting and promising solution, task offloading has been widespread concerned and is attracting great research attention. Along this direction, many research efforts have focused on mobile cloud computing, where mobile users can offload their computation-intensive tasks to the resource-rich remote clouds via wireless access  \cite{dinh2013survey}. Due to the labile wireless connection and high network latency between mobile devices and remote clouds, nevertheless, the satisfaction of the real-time interactive response requirements of mobile applications can be very challenging for such an approach. As an alternative, mobile edge computing is an emerging 5G service paradigm that leverages a multitude of collaborative end-user devices and/or near-user facilities at the network edge to carry out a substantial amount of communication and computation tasks \cite{hu2015mobile}. Since mobile edge computing is implemented in mobile users' close proximity, it can provide low-latency as well as agile computation and communication augmenting services for the users \cite{hu2015mobile}.

\begin{figure}[tt]
\centering
\includegraphics[height=2.0in]{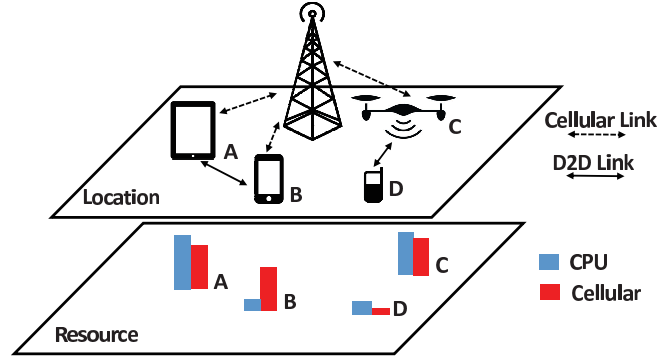}
\caption{An illustration of D2D Crowd framework. The dashed (solid) arrow represents cellular (D2D) link, and the blue (red) color bar represents available CPU (cellular bandwidth) resource.}
\label{fig:scenario}
\end{figure}

As illustrated in Fig. \ref{fig:scenario}, in this article we propose \emph{D2D Crowd}, a novel task offloading framework based on network-assisted Device-to-Device (D2D) collaboration, where a massive crowd of devices at the network edge can beneficially share the computation and communication resources among each other via the control assistance by the network operator. The common rationale is two-fold. On one hand, the operator  can typically have abundant network information for achieving more informed and efficient management decision making. On the other hand, the diverse capabilities of different types of devices (e.g., IoT devices, smartphones, and tablets) as well as the multiplexing gain (due to the runtime heterogeneity of resource availability among devices) can be exploited to support collaborative task execution for a variety of services. A simple illustration of this framework is shown in Fig. \ref{fig:scenario}. Suppose that a device $B$ would like to execute a computation-intensive task (e.g., data compression) while currently its CPU resource is heavily occupied by other applications. In this case, device $B$ can offload its task via the energy-efficient D2D link to a nearby device $A$, who possesses a large amount of idle CPU resource to facilitate the task execution.

We envision that, by jointly pooling and sharing heterogeneous computation and communication resources among the mobile devices, D2D Crowd can facilitate many novel applications and services demanding hybrid kinds of resources. For instance, the promising emerging applications enabled by the D2D Crowd framework include but are not limited to the following cases:
\begin{itemize}
\item \emph{Mobile Data Offloading}: Due to the heterogeneity of transmission technology adopted in device and the time-varying  nature of  wireless transmission, the quality of cellular connections among the devices can be diverse even at the same location. In this case, data offloading service can be carried out such that a device of poor cellular connection can offload its data to a nearby device with a high-quality cellular link in order to improve the energy efficiency.
\item \emph{Mobile Data Stream Processing}: Nowadays many mobile and IoT devices are equipped with a set of powerful embedded sensors and capable of acquiring and communicating a large amount of data streams. By leveraging a multitude of collaborative device computation resources at the network edge, D2D Crowd can enable efficient in-situ processing (e.g., data cleaning and feature abstraction) of the acquired data streams across variety of devices.
\item \emph{D2D-Assisted Cloud Offloading}: The D2D Crowd framework can also play a complementary role for the mobile cloud computing by provisioning the D2D-Assisted cloud offloading service. That is, instead of offloading the computation task to the cloud directly, a device (e.g., Device $D$ in Fig. \ref{fig:scenario}) can first transfer its computation task to a nearby device (e.g., Device $C$ in Fig. \ref{fig:scenario}) with both strong computing capability and good cellular connection, which can then help to process some small-scale task components and in parallel offload the computation-hungry components to the cloud\footnote{We will elaborate more on the integration of D2D-assisted cloud offloading with mobile-edge cloud computing service in Section \ref{MECC}.}.
\end{itemize}

Moreover, the D2D Crowd framework can achieve a win-win situation for both network operator and mobile users. On one hand, the users can benefit from the multiplexing gain due to the heterogeneity of resource availability among them. On the other hand, through good-quality cellular connection and various device resources sharing among users, the operator can extend its network coverage, improve its service quality as well as provide novel applications to attract more subscribers.

\subsection{Related Work}\label{RelatedWork}
Most existing researches on collaborative task offloading mainly focus on the delay tolerant networking (DTN) applications and the homogenous tasks utilizing the same type of resource (see \cite{shi2012serendipity,hu2014quality} and references therein).  For example, Shi \emph{et. al.}  in \cite{hu2014quality} design a traffic offloading framework to offload user's data files to nearby users. Hu \emph{et. al.} in \cite{shi2012serendipity} propose a task offloading framework where a mobile device offloads its CPU processing task to the encountered users.  Along a different line, in this article we consider the collaborative task offloading by heterogenous computation and communication resource sharing, which can enable many novel services in 5G networks. Moreover, motivated by the fact that the operator generally has sufficient network information and high computation power, we consider the \emph{network-assisted} architecture that the base station will provide the control assistance in determining the efficient task offloading assignments for mobile devices. Such a network assisted architecture is advocated by the emerging 5G networks (e.g., D2D overlaid cellular \cite{fodor2012design} and cellular IoT \cite{miorandi2012internet}) and software-defined wireless networks (e.g., SDN controller for application module scheduling \cite{pentikousis2013mobileflow}).

\subsection{Main Contributions}
In this article, we tackle the key issue of achieving energy-efficient task assignment, which is a critical build block of D2D Crowd, aiming at minimizing the total energy consumption for collaborative task executions among the devices. Specifically, we first introduce the D2D Crowd system model for joint computation and network resources sharing among the devices. Accordingly, we then propose the energy-efficient D2D Crowd task assignment problem formulation by accounting for the necessary constraints, and next develop a graph matching based optimal task assignment policy by leveraging the structural property of the problem. We further evaluate the performance of the proposed D2D Crowd task assignment policy through numerical study, which demonstrates a superior performance gain with more than 50\% energy consumption reduction over the case of local task executions. Last but not least, we also discuss the important directions of extending the D2D Crowd framework by taking into variety of application factors.

The rest of the article is organized as  follows. We first present the system model in Section \ref{TFramework}. We then propose the D2D Crowd task assignment problem formation and the graph matching based policy in Section \ref{TimeFrame}, and outline the directions of extending the framework in Section \ref{Further}. We finally conclude the article in Section \ref{cons}.

\section{D2D Crowd Model}\label{TFramework}
As illustrated in Fig. \ref{fig:scenario}, we consider that the D2D Crowd framework involves a set $\mathcal{N}=\{1,2,...,N\}$ of mobile devices and a network operator that manages multiple base stations. A device can establish the cellular connection with its associated base station (e.g., via LTE-Cat and LTE-M for smartphones and IoT devices, respectively) as well as the D2D connection with the devices in proximity (e.g., using cellular D2D or WiFi-direct).

\subsection{Device Resource Model}
We first introduce the model to describe the device resources of a D2D Crowd user for mobile task processing as follows.
\begin{itemize}
\item \emph{Computation Capacity}: For a device $i$, let $Z_i$ be its CPU working frequency, and hence the total computation capacity is $Z_i$ (in CPU cycles per unit time). In addition, we denote $\delta_i$ as its current load (i.e., percentage of occupied processing capacity) since the device $i$ may have the background load and run some unoffloadable tasks. Then, the available processing capacity for a D2D Crowd task is denoted by $c_i\!=\!(1\!-\delta_i)Z_i$.

\item \emph{Cellular Link}: Each device $i$ can establish a cellular link with its associated base station. We assume that the base station will determine a cellular transmission power level $P_i^b$ based on some power control scheme, and the device $i$ achieves an average cellular data rate $D_i$.

\item \emph{D2D Link}: Each device $i$ can also establish a D2D link with another device in proximity. The device can transmit data via both cellular and D2D links simultaneously  (i.e., using different transmission interfaces). We denote the D2D data rate from a user $i$ to another user $j$ as $D_{ij}$, and $P_i^d$ and $P_j^r$ are the D2D transmission power of device $i$ and the D2D receiving power of device $j$, respectively. As a global view from the network operator's perspective\footnote{Through network-assisted device discovery and local information report by the devices, the network operator can gather sufficient D2D connectivity information in practice.}, we introduce the D2D connectivity graph $G\!=\!\{\mathcal{N},\mathcal{E}\}$, where the set of devices $\mathcal{N}$ is the vertex set and $\mathcal{E}\!=\!\{(i,j)\!:\!e_{ij}\!=\!1,\ \forall i, j \in \mathcal{N}\}$ is the edge set where $e_{ij}\!=\!1$ if devices $i$ and $j$ can establish a feasible D2D link between them. Similar to FlashLinQ \cite{wu2013flashlinq}, due to the time and resource constraint we impose a practically-relevant constraint that a device can establish and maintain one D2D link during a task offloading round (e.g., a time frame).
\end{itemize}

\subsection{Mobile Task Model}\label{MobileTask}
In the D2D Crowd framework, we adopt a parameter tuple \emph{$<$$I_i, \varPsi_i, O_i, B_i$$>$} to characterize the mobile task of a device $i$, where $I_i$ is the input data size of the task, $\varPsi_i$ is the amount of computing resource required for the task (i.e., the number of CPU cycles), $O_i$ is the output data size of the task, and $B_i$ is the amount of cellular traffic required for the task. For simplicity, we do not distinguish between the upload and download cellular traffics, which can be easily factorized by introducing two separate cellular traffic parameters in the mobile task model.

We consider that this model allows rich service modeling flexibility in practice. For example, for the task of data uploading, we will have $\varPsi_i=0$ and $B_i$ as the cellular traffic to be uploaded. As another example, for the task of network-edge data stream processing, we have $\varPsi_i$ as the computing resource for processing the acquired data stream and $B_i$ as the required cellular traffic for transmitting the processed data to the remote data-center. Our model can be also easily extended to account for other types of resources (e.g., storage) by introducing more parameters in the tuple.

%In general, the task admission control and prioritization management can be carried out by a task queue controller in the upper layer (e.g., application layer) according to user specified need (e.g., the pull task queue in Google App Engine \cite{App_Engine}) or performance target (e.g., the push task queue in Google App Engine \cite{App_Engine}). Since we mainly focus on the network-assisted D2D collaboration for task offloading in this article, we simply consider a best-effort and first-come-first-serve task admission policy.

\subsection{Task Execution Model}\label{TaskExe}
We next introduce the task execution model such that a task can be either locally executed on its original mobile device or offloaded to be executed on another device in proximity.

\emph{1) Local Execution}: a device $i$  can locally execute its own task. According to the task parameter tuple, the execution time for computation is given by $T_i^c \!=\! \varPsi_i/c_i$, and the energy consumption is given by $E_i^c \!=\! \rho_i^cT_i^c$, where $\rho_i^c$ is the energy cost per CPU cycle for computation, which depends on the energy efficiency of the processor model and can be measured in practice \cite{kwakprocessor}. Similarly, the execution time for cellular communication is given by $T_i^b \!=\! B_i/D_i$, and the energy consumption is given by $E_i^b \!=\! P_i^bT_i^b$. Therefore, the total energy consumption of local execution is $E_i^l \!=\! E_i^c\!+\!E_i^b$.

\emph{2) Offloaded Execution}: As an alternative, a device $i$ can offload its own task to a nearby device $j$ via D2D link. In this case, the energy consumption for task input and output data transfer through D2D transmission between these two devices are given by $E_{ij}^d \!=\! (P_i^d+P_j^r)I_i/D_{ij}\!+\!(P_j^d+P_i^r)O_i/D_{ji}$. In addition, the energy overhead for executing the offloaded task in device $j$ is given by $E_{ij}^e \!=\! \rho_j^c\varPsi_i/c_j\!+\!P_j^bB_i/D_j$.
Therefore, the corresponding total energy consumption of offloaded execution is $E_{ij}^o \!=\! E_{ij}^d\!+\!E_{ij}^e$. Due to the constraint of physical size, mobile devices typically have limited resource capacity and hence we assume that a device can execute at most one task at a time.

\section{Energy-Efficient Task Assignment For D2D Crowd}\label{TimeFrame}
Based on the  system model above, we next describe the problem formulation for optimal task assignment in the D2D Crowd framework, in order to optimize the energy efficiency for collaborative task executions and meanwhile by taking into account the necessary assignment constraints.

\subsection{Problem Formulation}\label{SechdulingConst}

We first formulate the constraints for the D2D Crowd task assignment problem. Specifically, let $\pi_i$ be a binary indicator that is $1$ if device $i$ has a task to be executed, and $0$ otherwise (e.g., the task queue of device $i$ is empty). In addition, we adopt $\mu_{ij}$ as a binary decision variable for task assignment, which is $1$ if the task of a device $i$ is offloaded to execute on a device $j$, and $0$ otherwise. Note that $\mu_{ii}$ indicates whether a device $i$ locally executes its task.  For ease of notation, based on D2D connectivity graph $G\!=\!\{\mathcal{N},\mathcal{E}\}$, we also define that $e_{ii} \in \mathcal{E}$ (i.e., local execution is feasible). We then have the task assignment constraints as follow.
\begin{equation} \label{equ1}
\setlength{\abovedisplayskip}{3pt}
\setlength{\belowdisplayskip}{3pt}
\mu_{ij} \!=\!0, \quad \forall  e_{ij} \not \in \mathcal{E},
\end{equation}
\begin{equation} \label{equ2}
\setlength{\abovedisplayskip}{3pt}
\setlength{\belowdisplayskip}{3pt}
\sum\nolimits_{j \in \mathcal{N}}\mu_{ij}\!=\!\pi_i, \quad \forall  i \in \mathcal{N},
\end{equation}
\begin{equation} \label{equ4}
\setlength{\abovedisplayskip}{3pt}
\setlength{\belowdisplayskip}{3pt}
\sum\nolimits_{i \in \mathcal{N}} \mu_{ij} \!\leq\!1, \quad \forall  j \in \mathcal{N},
\end{equation}
\begin{equation} \label{equ6}
\setlength{\abovedisplayskip}{3pt}
\setlength{\belowdisplayskip}{3pt}
\mu_{ij} \!=\! \mu_{ji}, \quad \text{if} \ \pi_i \!=\! \pi_j \!=\! 1,
\end{equation}
\begin{equation} \label{equ3}
\setlength{\abovedisplayskip}{3pt}
\setlength{\belowdisplayskip}{3pt}
\mu_{ij} \!\in\! \{0,1\}.
\end{equation}
The constraint (\ref{equ1}) ensures that the task assignments are determined according to the feasible D2D connectivity.
The constraint (\ref{equ2}) represents that if a device has a task, its task should be assigned.
The constraint (\ref{equ4}) represents that due to the limited device resource, during a task offloading round a device will execute at most one task (which could either be its own task or an offloaded task from a nearby device)\footnote{For the case such that a device $i$ can execute multiple tasks simultaneously, we can extend the model by duplicating the device $i$ as multiple ``virtual" devices of the same resource capacity such that each virtual  device can execute at most one task.}.
\rev{The constraint (\ref{equ6}) presents that: (a) If two devices have tasks and they establish a D2D link between them to exchange their tasks with each other to execute, we have $\mu_{ij} \!=\! \mu_{ji}=1$. This is due to the fact that if a pair of devices have already established a D2D link during a task offloading round, they cannot establish any additional D2D link with other devices\footnote{As discussed in Section \ref{policy}, if a device can simultaneously establish multiple D2D links, then the task assignment problem is much easier, since it involves solving the standard bipartite matching problem.}; (b) If two devices have tasks and they do not want to establish a D2D link between them for task exchange, device $i$ (device $j$, similarly) can either execute the task locally or offload the task to another device $k$. In this case, we have $\mu_{ij} \!=\! \mu_{ji}=0$.}

By taking into account the assignment constraints above, our objective is to minimize the overall energy consumption of the task executions by all the devices, which is formally described as follows.
\begin{align} \label{equ11}
& \underset{\bm{\mu}}{\text{min}}
& & \sum\limits_{i=1}^N \left(\mu_{ii}E_i^l + (1-\mu_{ii})\sum\nolimits_{j \neq i}\mu_{ij}E_{ij}^o\right)\\
& \text{subject to}
& & (\ref{equ1}), (\ref{equ2}), (\ref{equ4}), (\ref{equ6}) \mbox{ and } (\ref{equ3}). \nonumber
\end{align}

\begin{figure}[tt]
\centering
\includegraphics[scale=1.0]{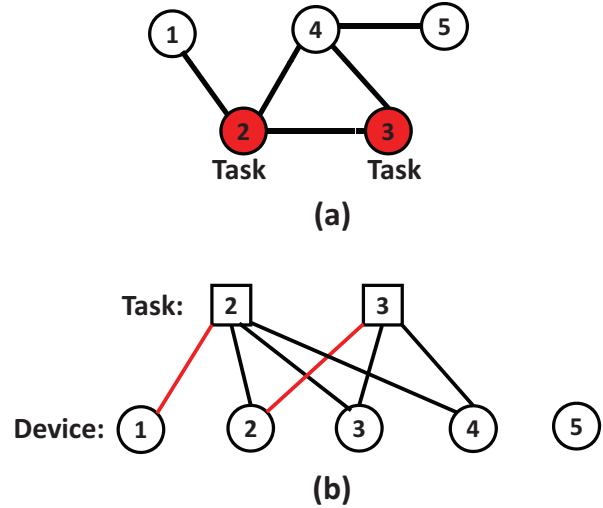}
\caption{An illustration of bipartite matching based task assignment, with (a) the D2D connectivity graph and (b) the transformed bipartite matching.}
\label{fig:BMatching}
\end{figure}

\begin{figure}[tt]
\centering
\includegraphics[scale=0.8]{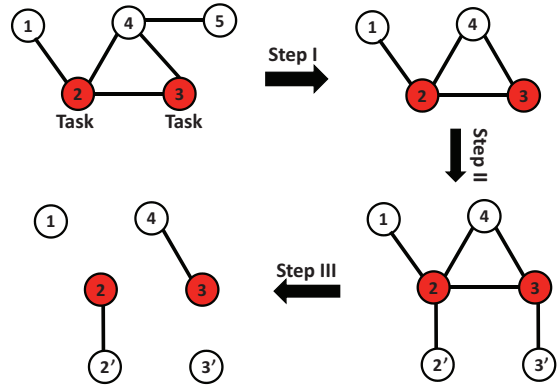}
\caption{An illustration of the modified graph matching based task assignment, with pruning in Step I, replication in Step II and matching in Step III.}
\label{fig:GMatching}
\end{figure}

\subsection{Graph Matching Based Optimal Task Assignment}\label{policy}
We next propose the optimal task assignment policy for the problem in (\ref{equ11}).

At a first glimpse, one might regard problem (\ref{equ11}) as the classical assignment problem, which can be solved by the minimum weight bipartite matching solution. That is, as illustrated in Fig. \ref{fig:BMatching},  a weighted bipartite graph can be constructed, where on one side a node $i$ represents a task of a device $i$, and on the other side the nodes represents the set of devices for the task executions. There exists an edge between a task node $i$ and a device node $j$ if there is a feasible D2D link between devices $i$ and $j$, and the edge weight denotes the energy consumption. Then, the minimum weight bipartite matching solution can be found using the standard Hungarian algorithm \cite{even2011graph}.

However, the above approach would fail to work in our problem. Taking Fig. \ref{fig:BMatching} for instance, suppose the minimum weight bipartite matching solution is that device $2$'s task is assigned to device $1$ and device $3$'s task is assigned to device $2$. In this case, device $2$ needs to simultaneously establish and maintain two D2D links with devices $1$ and $3$, which would demand excessive resource and overhead and hence violates the assignment constraint in (\ref{equ6}).

Thus, in this article we propose a novel graph matching based solution for problem (\ref{equ11}). The most critical part is that we need to properly define the graph structure for matching. The key idea of our solution is that we first adopt the D2D connectivity graph $G$ as the basic graph to ensure the structure of assignment feasibility via D2D links. Then we modify the basic graph using the following two steps: 1) \emph{Node Pruning}: for a device $k$ that does not have a task, if all its neighboring devices also do not have any tasks, then we remove the device node $k$ in the graph (e.g., node $5$ is removed in Step I in Fig. \ref{fig:GMatching}), since the device $k$ will not be assigned with any task; 2) \emph{Node Replication}: for a device node $i$ that has a task to be executed (i.e., $\pi_{i}=1$), we add a new replicated node $i'$ and add an edge connecting node $i'$ and node $i$ (e.g., nodes $2$ and $3$ have tasks and hence two replicated nodes are added in Step II in Fig. \ref{fig:GMatching}). By doing so, for the matching over the modified graph, we have the following three nice representations:
\begin{itemize}
\item Case 1: If node $i$ is matched with its replicated node $i'$, then it means that the device $i$ will execute its task locally.
\item Case 2: If node $i$ is matched with another node $j$ without a replicated node, then it represents that the task of device $i$ will be offloaded to device $j$.
\item Case 3: If node $i$ is matched with another node $j$ having a replicated node, then it indicates that devices $i$ and $j$ exchange their tasks for the mutually offloaded execution\footnote{For example, two devices have heterogenous resource capacity and they are running tasks with different resource requests. Mutual task offloading can be beneficial.}.
\end{itemize}

Moreover, due to the property of graph matching, each device node will be matched to at most one node. This ensures the assignment constraint in (\ref{equ6}) can be satisfied. Accordingly, we next define the weights for the edges of the modified graph as follows:
\begin{itemize}
\item Case 1: for an edge that connects node $i$ with its replicated node $i'$, we set the weight $w_{ii'}=E_i^l$, i.e., the energy consumption of local execution.
\item Case 2: for an edge that connects node $i$ with another node $j$ without a replicated node, we set the weight $w_{ij}=E_{ij}^o $, i.e., the energy consumption of offloaded execution from device $i$ to device $j$.
\item Case 3: for an edge that connects node $i$ with another node $j$ having a replicated node, we set the weight $w_{ij}=E_{ij}^o+E_{ji}^o$, i.e., the total energy consumption of the mutually offloaded executions. \rev{In this case, devices $i$ and $j$ would exchange their tasks for execution and hence the edge weight should represent the total energy consumption of these two devices.}
\end{itemize}

Based on the modified weighted graph above, we can obtain the optimal solution for the D2D Crowd task assignment problem by finding the minimum weight matching solution using the Edmonds's Blossom algorithm \cite{even2011graph}, which possesses a polynomial time complexity and can compute the solution in a fast manner as shown in the performance evaluation section later. As illustrative example, we show a matching solution in Step III in Fig. \ref{fig:GMatching} where node $2$ is matched with  node $2'$ (i.e., device $2$ executes the task locally) and node $3$ is matched with node $4$ (i.e., device $3$ offloads its task an idle device $4$).

\begin{table}[!htbp]
\small \centering
\caption{Simulation Setting}\label{tab:paraTable}
\begin{tabular}{|c|c|c|}
\hline
Category & Parameter& Value\\
\hline
\multirow{2}*{Cellular Link}
                   &Cellular Data Rate &[1, 10]\,Mbps \\
                   &Cellular Power &600\,mW\\
\hline
\multirow{5}*{D2D Link}
                   &Maximum D2D Bandwidth  &20\,Mhz \\
                   &Maximum D2D Distance  &200\,m \\
                   &D2D Power &200\,mW\\
                   &Path Loss Exponent & 3\\
                   &Background Noise & $10^{-8}$\\
\hline
\multirow{6}*{User \& Task}
                   &Total CPU Cycles &2\,Ghz\\
                   &CPU Load &[0, 70\%]\\
                   &CPU Power &900\,mW\\
                   & Task Input Size & [500, 2000]\,KB \\
                   & Processing Density of Pure CPU Tasks &3000 cycle/bit \\
                   & Processing Density of Hybrid Tasks &1000 cycle/bit \\

\hline
%\multirow{4}*{Task}
%                   & Generation Frequency (Poisson) &0.4 \\
%%                   &CPU (Cellular) Type Probability &50\% \\
%
%\hline
%\multirow{3}*{Incentive}
%                   &$\alpha^c, \alpha^d, \alpha^b$ For All Users  &0.5 \\
%                   &$\beta^c, \beta^d, \beta^b$ For All Users  &0.01 \\
%                   &$K^o$ For All Users & 0.25\,J \\
%%                   &Control Parameter $V$ & 30 \\
%
%%                   &Simulation Period  &$10000$\,s \\
% \hline
\end{tabular}

\end{table}

\subsection{Performance Evaluation}\label{evaluation}
We next evaluate the performance of the proposed  task assignment policy for D2D Crowd through numerical studies. Table \ref{tab:paraTable} shows the simulation parameter settings, most of which are in accordance with the real measurements in practice \cite{kwak2015dream}. In the simulation, we run 100 rounds of task assignments to obtain the average performance. The feasible D2D connectivity among the devices is varying from round to round, which depends on the devices' locations. Here we use the commonly-adopted Opportunistic Network Environment (ONE) simulator\cite{ekman2008working} to model devices' location dynamics due to mobility, which has been shown to well capture the distributional property in many real-world user mobility traces \cite{ekman2008working}.

We consider three different types of application tasks, including pure CPU tasks (e.g., image compression), pure cellular tasks (e.g., file downloading) and hybrid tasks requiring both CPU and cellular resources (e.g., data stream acquisition and analytic). In the simulation we adopt the data processing applications as the study case, such that the required computation cycles for task execution is proportional to its input data size (described by the processing density in Table \ref{tab:paraTable}). During each task assignment round, a device's task is generated according to the task generation frequency, i.e., the probability that a device has a task. We use the energy saving ratio as the performance metric which is defined as the energy consumption of a task assignment scheme compared with the result that each device executes its own task locally.

\begin{figure}[tt]
\centering
\includegraphics[scale=0.35]{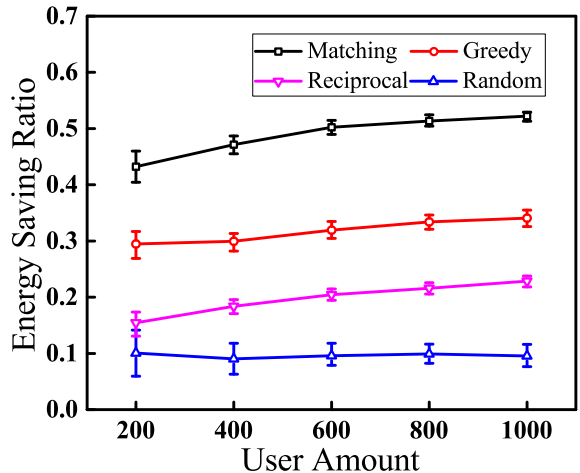}
\caption{Energy saving ratios with different number of user devices}
\label{fig1}
\end{figure}

\begin{figure}[tt]
\centering
\includegraphics[scale=0.35]{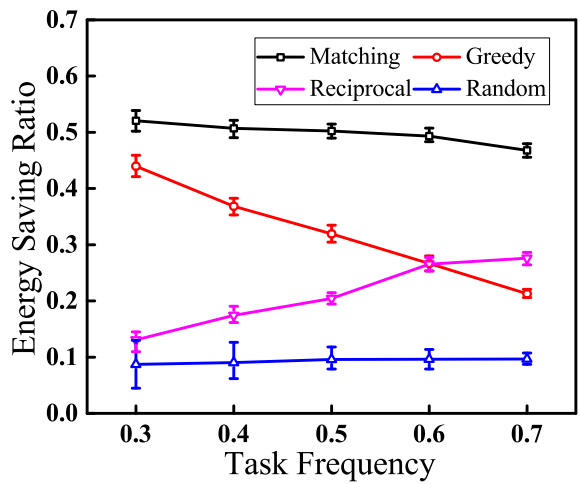}
\caption{Energy saving ratios with different task generation frequencies}
\label{fig2}
\end{figure}

\begin{figure}[tt]
\centering
\includegraphics[scale=0.35]{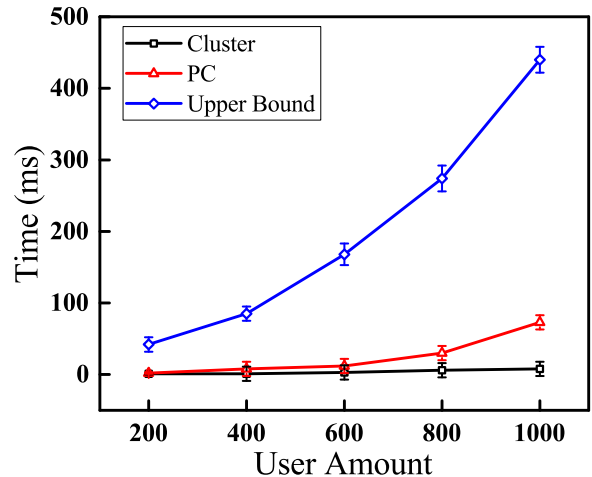}
\caption{Running time}
\label{fig3}
\end{figure}

W compare our graph matching based task assignment policy with three schemes:
1) \emph{Greedy}: the base station first sorts all feasible task owner-executor pairs, and then chooses the owner-worker pairs greedily;
2) \emph{Reciprocal}: two users are reciprocal if both of them are task owners and achieves mutually better performance when they exchange the tasks for execution. In this scheme, the base station only selects the pairs of users who are reciprocal;
3) \emph{Random}: the base station first randomly sorts the task owners, who will act sequentially to choose a feasible neighboring device randomly among the remaining devices for task execution.

We depict the energy saving ratio by different schemes with different user amounts in Fig. \ref{fig1}, in which we also plot the error bar with 90\% confidence interval. We see that our proposed graph matching based task assignment scheme can achieve an high energy saving ratio of 50\%.  Also, our scheme can save more than 40\%, 30\%, and 20\% energy compared with the random, reciprocal, and greedy schemes, respectively. This demonstrates the superior performance of our task assignment scheme for D2D Crowd. We observe that the performance of our algorithm will slightly increase with user amount increasing. The reason is that larger user size would enable a device to have more neighbors, and hence a device can have more opportunities to offload the task to a suitable device.

To evaluate the impact of task amount, we then show the energy saving ratio by different schemes with different task generation frequencies in Fig. \ref{fig2}. We observe that with task frequency increasing both the performance of our approach and the greedy scheme decreases (greedy scheme has a more significant performance drop, while our approach is more robust and slightly decreases). This is due to in large part that as task frequency increases, more devices will generate tasks. This will decrease the number of idle devices, and hence hinders the overall performance. In addition, with more user generating task the task reciprocal pairs are also increasing, which makes the performance of reciprocal scheme improve. Despite that, our scheme can still save more than 39\%, 28\% and 20\% energy over random, greedy and reciprocal schemes, respectively, even when the task generation frequency is high.

We next evaluate the running time of the proposed task assignment scheme based on Edmonds's Blossom algorithm. We run our scheme on an ordinal computer with  Intel Core i5-2400 Processor@3.1GHz and 8G Memory.  Fig. \ref{fig3} shows the average running time in terms of million-seconds (ms).
\rev{It shows that the running time almost linearly (super-linearly) increases as the number of devices increase. The running time is much less than the theoretic upper bound of Edmonds's Blossom algorithm, less than 80ms even when the number of devices is large (e.g., 1000). This demonstrates our scheme can obtain the task assignment solution in a fast manner. Furthermore, we implement in our scheme on a computing cluster (Dell PowerEdge C6100), which is generally deployed at the network edge by the operator for scheduling management.  We observe a very fast running time of less than 1 ms in all cases, which corroborates our proposed scheme is amenable for practical application.}

\section{Extensions and Future Directions}\label{Further}
In the sections above, we mainly introduce the D2D Crowd framework and focus on addressing the key issue of energy-efficient collaborative task assignment. In this part, we will further discuss the several important directions for extending D2D Crowd into a full-fledged framework by accounting for variety of application factors.

\subsection{Integration with Mobile-Edge Cloud Service}\label{MECC}
As an emerging service for mobile edge computing, mobile edge cloud computing is a paradigm to provide augmented cloud computing capabilities at the edge of pervasive radio access networks in close proximity to mobile devices \cite{hu2015mobile}.

As mentioned earlier, the proposed D2D Crowd framework can be well integrated with the mobile edge cloud computing to further boost the performance gain of cloud offloading. For instance, for a device $i$ that would like to utilize the mobile-edge cloud service, we can define two execution modes: 1) \emph{Direct Cloud Offloading}: device $i$ utilizes its own cellular link to offload its computation task to the mobile edge cloud at the bast station directly; 2) \emph{D2D-Assisted Cloud Offloading}: using the energy-efficient D2D link, device $i$ first transfers its computation task to a nearby device $j$ of a good cellular connection, and then device $j$ helps to offload  device $i$'s task to the mobile-edge cloud. Furthermore, when device $j$ also has a strong computing capability, device $i$ can also decide to partition a portion of its computation task to be executed at device $j$, to reduce the offloading volume to the mobile-edge cloud and the energy overhead as well. For such an integrated computing service framework, an important emerging research issue is that how to properly partition the computation task components among the local device, the D2D offloaded device and the mobile-edge cloud in order to optimize the overall energy efficiency.

\subsection{Incentive Mechanisms for Collaboration}
For practical implementation, the D2D Crowd framework strongly relies on device's collaboration, and hence a
good incentive mechanism that can prevent the over-exploiting and free-riding behaviors that harm device user's
motivation for collaboration is highly desirable. As an initial attempt, we next discuss a network-assisted incentive mechanism tailored to the D2D Crowd framework as follows.

The incentive mechanism is motivated by the resource tit-for-tat scheme in peer-to-peer systems. The key idea is to ensure that a device is allowable to exploit more resources from other devices only if it has contributed sufficient resources to the others. We can regard the resource contribution as user credit, which will be maintained by the base station.

Specifically, we denote by $X_i^{CPU}(t)$ and $X_i^{Cellular}(t)$ to represent the amount of CPU resource (e.g., in terms of CPU cycles) and the cellular resource (e.g., in terms of data bytes) that other users contribute to a device $i$, respectively, and $Y_i^{CPU}(t)$ and $Y_i^{Cellular}(t)$ to represent the resource amounts that the device $i$ contributes to other devices up to time $t$. We then define the following resource tit-for-tat constraint
\begin{align}
& \alpha_i^{CPU} X_i^{CPU} \leq \beta_i^{CPU}+Y_i^{CPU}, \label{equ7}\\
& \alpha_i^{Cellular} X_i^{Cellular} \leq \beta_i^{Cellular}+Y_i^{Cellular},  \label{equ8}
\end{align}
where the parameters $\alpha_i^*$ and $\beta_i^*$ are normalized within the range [0,1]. The resource tit-for-tat constraint reflects that the resource amount that a device exploits from the others is in proportion to that it contributes to the others. If a device wants more resources, it needs to share more in return. To promote the collaboration and resource contribution, when a device does not satisfy resource tit-for-tat constraints above, the network operator will not assign device $i$'s offloaded tasks.

\subsection{Coping with System Dynamics}
In order to gain useful insights, in the discussions above we mainly discuss the D2D Crowd framework  in the static setting and consider the task assignment issue during each task offloading round. To implement the proposed D2D Crowd framework in practical systems, we need to consider its generalization to deal with the system dynamics. In a dynamic setting, many system factors are time varying, e.g., devices' D2D connections can dynamically change due to device users' mobility, and the cellular link quality can vary from time to time due to fading effect.  Moreover, different number of new tasks can be generated at a device across different time frames, and hence we need to carefully address the task queueing issues, to prevent the queue explosion from a long run perspective.

To cope with the system dynamics, we can consider to generalize the current D2D Crowd task assignment scheme by resorting to the tool of Lyapunov optimization \cite{neely2010stochastic}.  Generally speaking, two key salient features enable Lyapunov optimization suitable for addressing dynamic task assignment problem: 1) Lyapunov optimization is an online stochastic optimization problems with time-average objective, and in general it only utilizes the system information at the current time period; 2) It also enables the feature of stabilizing the queues by providing a drift-plus-penalty function for joint queue stability and time-average objective optimization. Thus, in a future work we can explore to leverage the Lyapunov optimization approach to design efficient online task assignment for D2D Crowd that can be adaptive to the system dynamics and meanwhile can ensure the stability of the task queues.
%
%We should emphasize that the proposed graph matching based task assignment policy in Section \ref{policy} can lay down a solid foundation for designing the online task assignment mechanisms. This is due to that the the drift-plus-penalty function optimization problem in the Lyapunov optimization approach typically shares similar structure property with the original problem without the penalty term for queue stability.  For instance, the weights in the modified graph for task assignment can be different when accounting for the penalty for queue stability, but the key principle of the proposed graph modification and matching can apply.

\subsection{Hybrid Centralized-Decentralized Implementation}
A key focus in this article is to embrace the benefit of network assisted D2D collaboration for energy-efficient mobile edge computing. The network assisted  architecture enables an efficient centralized management paradigm, and hence is advocated in many future networking system, e.g., 5G D2D overlaid cellular networks, cellular IoT, and software-defined mobile networks.

Another important direction of further extending the proposed D2D Crowd framework is to consider its hybrid centralized-decentralized implementation. This can be highly relevant to the application scenario that we would like to achieve a synergetic scheduling across multiple heterogenous networks such as cellular network and WiFi network. The decentralized nature of CSMA access in WiFi requires a hybrid centralized-decentralized design. Also, in some cases some gateway device can have better communications links with its peripheral devices and hence can be selected as a leader device for achieving efficient local coordination.

One possible approach for implementing the hybrid centralized-decentralized paradigm is that we can first decompose the D2D connectivity graph into multiple  communities. Within a community, one leader device can be selected or elected to manage the local task assignment (e.g., using the proposed graph matching based scheme locally) for D2D collaboration based mobile edge computing. The leader devices will also negotiate among themselves for task assignment synchronization and conflict resolution. How to design the lightweight and efficient protocol for such a hybrid centralized-decentralized implementation can be very interesting and challenging.

\section{Conclusion}\label{cons}
\rev{In this article, we proposed a novel framework of D2D Crowd, as an emerging key service for 5G mobile edge computing, by leveraging a massive crowd of devices at the network edge for energy-efficient collaborative task executions. By jointly pooling and sharing heterogeneous computation and communication resources among the mobile devices, D2D Crowd can facilitate many novel applications and services demanding hybrid kinds of resources.

Specifically, we proposed the system model for D2D Crowd and formulated the energy-efficient D2D Crowd task assignment problem. We further proposed a graph matching based optimal task assignment policy, and showed that it can  achieve a superior performance through numerical evaluation. Finally, we discussed several important and interesting directions for extending the D2D Crowd framework, such as devising a proper incentive mechanism for encouraging device collaboration and developing an efficient online mechanism for coping with the system dynamics.}

%\section*{Conclusion}

\bibliographystyle{IEEEtran}
\bibliography{Fog}

\end{document}